\documentclass[letterpaper]{article} 
\usepackage{aaai24}  
\usepackage{times}  
\usepackage{helvet}  
\usepackage{courier}  
\usepackage[hyphens]{url}  
\usepackage{graphicx} 
\usepackage{natbib}  
\usepackage{caption} 
\newcommand{\cmark}{\ding{51}}%
\newcommand{\xmark}{\ding{55}}%
\newcommand{\qmark}{\textbf{\textit{?}}}
\usepackage{algorithm}
\usepackage{algorithmic}
\usepackage{cleveref}
\usepackage{pifont}
\newcommand{\xhdr}[1]{\vspace{1.5ex}\noindent\textbf{#1}}

\usepackage{listings}
\usepackage{xcolor}
\usepackage{booktabs}
\usepackage{lipsum}
\definecolor{codegreen}{rgb}{0,0.6,0}
\definecolor{codegray}{rgb}{0.5,0.5,0.5}
\definecolor{codepurple}{rgb}{0.58,0,0.82}
\definecolor{backcolour}{rgb}{0.95,0.95,0.92}

\lstdefinestyle{mystyle}{
    backgroundcolor=\color{backcolour},   
    commentstyle=\color{codegreen},
    keywordstyle=\color{magenta},
    numberstyle=\tiny\color{codegray},
    numbers=none,
    stringstyle=\color{codepurple},
    basicstyle=\ttfamily\scriptsize,
    breakatwhitespace=false,         
    breaklines=true,                 
    captionpos=b,                    
    keepspaces=true,                 
    numbers=left,                    
    numbersep=5pt,                  
    showspaces=false,                
    showstringspaces=false,
    showtabs=false,                  
    tabsize=2
}
\newcommand{\lstcolourline}[1]{\rlap{\color{#1}\rule[-.3\baselineskip]{\linewidth}{\baselineskip}}}

\DeclareRobustCommand{\recwizard}{%
  \begingroup\normalfont
  \raisebox{-0.12em}{%
  \includegraphics[height=1.1em]{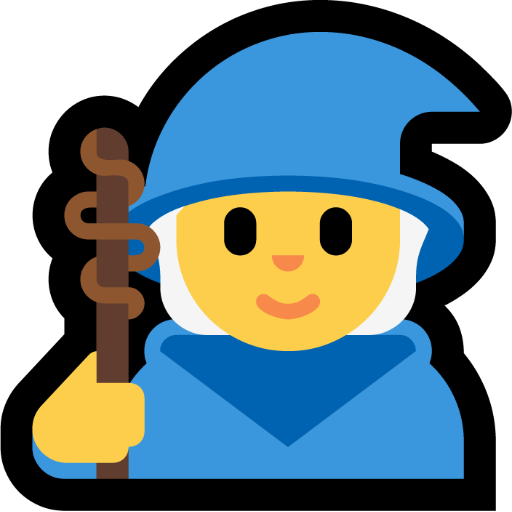}%
  }%
  \kern 0.4em%
  \endgroup
}

\title{\recwizard \texttt{RecWizard}: A Toolkit for Conversational Recommendation with \\ Modular, Portable Models and Interactive User Interface}

\author {
    Zeyuan Zhang\equalcontrib,
    Tanmay Laud\equalcontrib,
    Zihang He\equalcontrib,
    Xiaojie Chen,
    Xinshuang Liu,
    Zhouhang Xie,
    Julian McAuley,
    Zhankui He
}
\affiliations {
    University of California, San Diego
\\
    {\{zez018,
    tlaud,
    z6he,
    xic051,
    xil235,
    zhx022,
    jmcauley,
    zhh004\}@ucsd.edu}
}

\usepackage{bibentry}

\begin{document}

\maketitle

\begin{abstract}

We present a new Python toolkit called \textbf{\texttt{RecWizard}} for Conversational Recommender Systems (CRS). \textbf{\texttt{RecWizard}} offers support for development of models and interactive user interface, drawing from the best practices of the Huggingface ecosystems. CRS with \textbf{\texttt{RecWizard}} are \emph{modular, portable, interactive} and \emph{Large Language Models (LLMs)-friendly}, to streamline the learning process and reduce the additional effort for CRS research. For more comprehensive information about \textbf{\texttt{RecWizard}}, please check our GitHub \texttt{\url{https://github.com/McAuley-Lab/RecWizard}}.

\end{abstract}

\section{Introduction}

Conversational Recommender Systems (CRS)~\citep{christakopoulou2016towards,li2018towards} are gaining increasing attention from industry and academia, especially with the emergence of Large Language Models (LLMs)~\citep{friedman2023leveraging,he2023large,wang2023rethinking}. To expedite CRS research, there is a pressing need for an open-source toolkit that lowers the barrier to reusing CRS and LLMs resources, developing new CRS, and interacting via user interface for debugging or evaluation. However, current toolkits fail to meet such requirements~\citep{miller2017parlai,zhou2021crslab,quan2022force}.


To address these limitations, we propose \textbf{\texttt{RecWizard}}, a Hugging Face~\citep{wolf2020transformers} (HF)-based CRS toolkit for research purposes, with the following properties:
\begin{itemize}
    \item \textbf{Modular}: We abstract CRS to a lower \textit{module level} and a higher \textit{pipeline level} and make them modular.
    \item \textbf{Portable}: With HF compatibility, modules and pipelines can be effortlessly shared online and easily deployed.
    \item \textbf{Interactive}: A user-friendly interactive interface is provided for conversations between users and CRS.
    \item \textbf{LLMs-Friendly}: We demonstrate LLMs as different roles, e.g., recommender modules, in RecWizard.
\end{itemize}

\section{Related Works}

\begin{table}[]
\small
\centering
\begin{tabular}{lccc}
\toprule
\textbf{}                & \textbf{CRSLab} & \textbf{FORCE} & \textbf{Ours.} \\ \midrule
\textbf{Open-Source?}    & \cmark               & \xmark                   & \cmark              \\
\textbf{Modular?}        & \qmark               & \cmark              & \cmark     \\
\textbf{Portable?}       & \xmark             &  \xmark                  & \cmark              \\
\textbf{User Interface?} & \xmark               & \cmark              & \cmark              \\ 
\textbf{LLM-Supported?} & \xmark               &  \xmark                   & \cmark              \\\midrule
\textbf{CRS Setting}         & Flexible            & Rule-based     & Flexible           \\  \bottomrule
\end{tabular}
\caption{The properties of \textbf{\texttt{RecWizard}}~(ours.) compared to two previous CRS related toolkits, CRSLab~\citep{zhou2021crslab} and FORCE~\cite{quan2022force}} 
\label{tab:comp}
\end{table}

Traditional recommender-system toolkits~\citep{zhao2021recbole,zhao2022recbole,ivchenko2022torchrec} only support users' actions (e.g., clicks or purchases), not CRS, and general-purpose conversational-system toolkits~\citep{miller2017parlai} are not designed for CRS use cases.
We outline the existing CRS toolkits~\citep{zhou2021crslab, quan2022force} and highlight the new properties in our RecWizard in Table~\ref{tab:comp}. Detailed explanations are provided in our supplementary.

\section{Design Principles} 
\begin{figure*}[htbp]
  \centering
    \includegraphics[width=\textwidth]{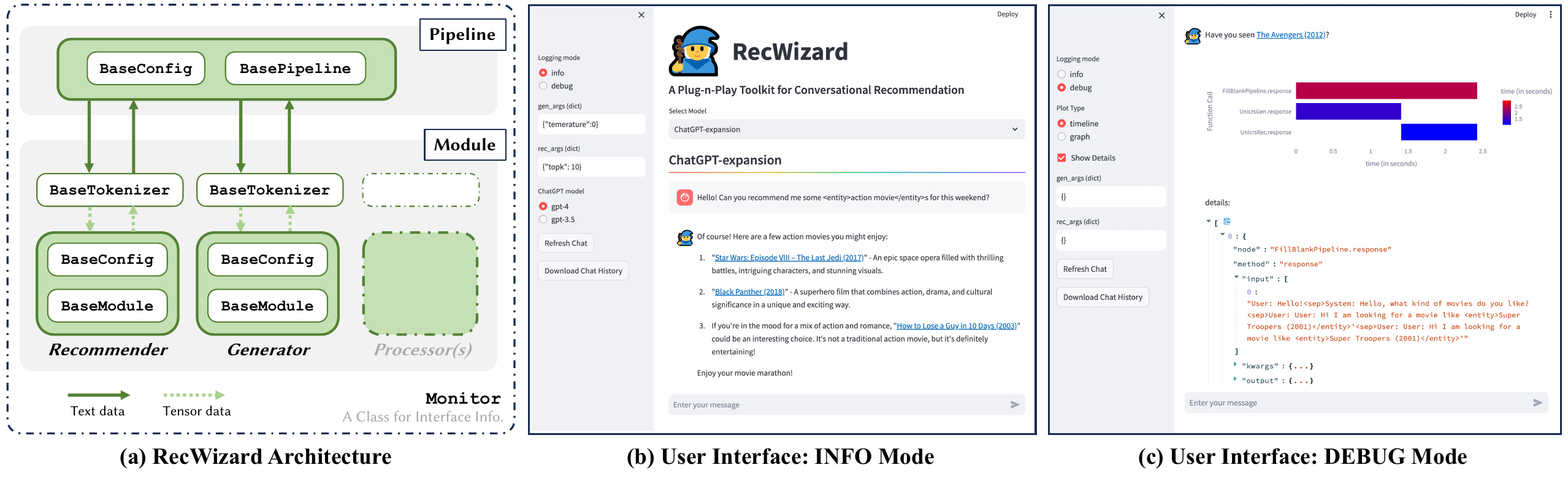}
  \caption{\textbf{(a)} The RecWizard architecture comprises pipeline and module levels. Text data flows between modules and the pipeline, while tensor data flows within modules after being processed by RecWizard tokenizers, ensuring RecWizard modularity and portability. \textbf{(b)} In INFO mode, the RecWizard example \emph{ChatGPT-expansion} includes the ReDIAL-Rec recommender module and ChatGPT with prompts as the generator module. Users can select models, set basic arguments, and chat with RecWizard. \textbf{(c)} In DEBUG mode, using the \emph{UniCRS-ReDIAL} model as an example, module-level timeline visualization and intermediate messages are enabled for debugging or in-depth demonstrations in the user interface.}
  \label{fig:main}
\end{figure*}

We design an abstraction for CRS with RecWizard at two levels, as shown in Figure~\ref{fig:main} (a):
\begin{enumerate}
    \item \textbf{Module level:} We offer a \textit{recommender} module for recommendations and a \textit{generator} module for natural-language responses, as many papers implemented~\cite{li2018towards,chen2019towards,wang2022towards}. Further, \textit{processor} modules can be integrated to extract vital information, e.g., via \textit{entity linking}, from users' raw text.
    \item \textbf{Pipeline level:} A RecWizard pipeline is a high-level logic that determines when and how to call the modules and how to aggregate the results from these modules.
\end{enumerate}
By default, all modules communicate using natural language (\textbf{text data}), ensuring maximum modularity. However, we also provide developers with the lower-level methods to define module communication through \textbf{tensor data} in a flexible and differentiable manner.

\section{Why Use RecWizard}
In this section, we present RecWizard use-cases for flexible \emph{development} and deploying \emph{interactive user interface}.

\subsection{Effortless Usage} 
\label{sec:deloy}
\xhdr{Module Level:} CRS practitioners can easily load any trained recommender (or generator) modules from the HF hub. For example, a UniCRS~\citep{wang2022towards} recommender module pretrained on ReDIAL dataset can be loaded seamlessly ~\citep{li2018towards}:

\begin{lstlisting}[language=python]
  unicrs_rec = recwizard.UnicrsRec \ 
            .from_pretrained('unicrs-rec-redial')
\end{lstlisting}

\xhdr{Pipeline Level:} To build a RecWizard model, the next step is loading a high-level pipeline. Apart from loading a pipeline with a one-liner similar to the module-level example above, it is also flexible to create UniCRS model variants by combining different modules under a certain pipeline. Here we use a new ChatGPT~\citep{openai2022chatgpt}-based generator module to build a \textbf{\texttt{unicrs\_variant}}:

\begin{lstlisting}[language=python]
  unicrs_variant = recwizard.FillblankPipeline(
      config=config, 
      rec_module=unicrs_rec, 
@\lstcolourline{yellow!15}@      gen_module=recwizard.ChatGPTGen \
@\lstcolourline{yellow!15}@        .from_pretrained('chatgpt-gen-fillblank'))
\end{lstlisting}

\noindent This example highlights the modular, portable, and LLM-friendly design of RecWizard, which simplifies the usage of the trained modules or pipelines. Detailed examples are provided in our supplementary.

\subsection{Flexible Development} 
RecWizard Tokenizers are the key to our framework, bridging the text interface between modules. At its core, it uses the ``composite pattern'' to extend HF tokenizers for CRS. Further, it provides a framework to parse information (like entities) in text-only format that we designed for conversational recommendation (see supplementary).

New modules can be specified by defining the configuration, tokenizer, and module classes as below\footnote{We omit the configuration and tokenizer classes due to limited spacing. Please see the complete code template in supplementary.}: 

\begin{lstlisting}[language=python]
  class NewModule(recwizard.BaseModule):
        # for recommender, generator or more processors
      def __init__(self, ...):
        # build it similar to HF PreTrainedModel
      def forward(self, tensor_inputs, labels):
        # define the flow of tensors through the module
      def response(self, raw_inputs, tokenizer):
        # define inputs and outputs based on forward
\end{lstlisting} 

\noindent Pipelines are specified by configuration and \textbf{\texttt{response}} method that defines the execution flow of the modules. 
All our Base classes are designed to be compatible with HF's \textbf{\texttt{push\_to\_hub}} method (see more details in supplementary).

\subsection{User-Friendly Interface} We provide an easy-to-use user interface for our models to support two different control modes in Figure~\ref{fig:main} (b)-(c):

\xhdr{INFO Mode}: Users can chat with a selected system, asking for natural language responses or recommended items. We suggest using this mode in cases such as : (1) demonstrating the final RecWizard model; (2) inviting users for human evaluation.

\xhdr{DEBUG Mode}: Users observe the module execution timeline, intermediate results, and control the internal module arguments. We suggest this mode for (1) debugging the pipeline at the module level and (2) understanding or explaining how a certain RecWizard pipeline works.

\section{Conclusion and Future Work}

We present \textbf{\texttt{RecWizard}}, a toolkit for Conversational Recommender Systems (CRS) research based on Hugging Face. RecWizard offers user-friendly deployment and development APIs and user interface for conversation interaction and debugging. We plan to share more pipelines and trained modules for CRS research and contribute to CRS benchmarking as well as online services in the future.

\bibliography{aaai24}

\newpage

\appendix 

\section{Discussion on CRS Toolkits}
\xhdr{On CRSLab~\citep{zhou2021crslab}.} Regarding CRSLab~\citep{zhou2021crslab}, it is widely used in CRS research as a comprehensive Python library for reproducing a collection of popular CRS model results from published papers. However, there are some limitations that we should consider:
\begin{enumerate}
    \item CRSLab mainly focuses on recommendation or generation quality at the \textbf{module level} only because the main results reported in related publications are usually at the module level. However, focusing solely on separate CRS modules while ignoring their evaluation as a whole can lead to various issues, such as recommendation-generation inconsistencies in final responses~\citep{wang2022towards} and difficulties in detecting trivial shortcuts in responses~\citep{he2023large}. We hope RecWizard, with its emphasis on the \textbf{pipeline level}, can draw more attention in the CRS community to research problems at this level.
    \item CRSLab does not offer an interactive user interface for CRS. Since CRSLab primarily focuses on the performance of individual modules, it lacks the necessary tools to integrate these modules into a cohesive whole and subsequently deploy interactive user interface for CRS models. RecWizard provides an easy-to-use interface, and we believe this feature will be helpful for a more realistic and comprehensive CRS evaluation in the future.
    \item CRSLab is built upon the native PyTorch framework, which is well-organized. However, with recent rapid developments, e.g., in LLMs, the Hugging Face (HF) framework, backed by PyTorch, offers more research-friendly features: \textbf{(1) Portability:} Users or developers can easily share datasets or trained models online and load them with minimal code. \textbf{(2) LLM-friendliness:} Many LLMs are released under the HF framework, making them readily available for reuse. Therefore, compared to CRSLab, RecWizard inherits these features from HF and adapts them for CRS research, promoting more user-friendly deployment and development.
\end{enumerate}

\xhdr{On FORCE~\citep{quan2022force}:} FORCE~\citep{quan2022force} is an in-house tool at Microsoft and is published as a conversational recommendation demo. FORCE offers a well-designed user interface, but it has some limitations, like:

\begin{enumerate}
    \item FORCE is not available as an open-source toolkit. Therefore, the research community cannot access the CRS user interface and its associated features, which limits the potential impact of FORCE.
    \item FORCE offers limited CRS settings. In the paper and video description, the authors exclusively present a Rule-based setting, where the system must determine ``question/chat/recommendation'' actions before responding, and a knowledge graph is mandatory. This requirement significantly restricts the flexibility for research purposes compared to CRSLab or our RecWizard.
    \item FORCE aims to visualize and facilitate the data collection process for Rule-based CRS models rather than providing a portable or LLMs-friendly toolkit for CRS research, such as new model development and sharing.
\end{enumerate}

\section{Current CRS Models in RecWizard}

Table \ref{tab:resource} shows the current CRS models in RecWizard. We first build them at the module level, following the module setup listed in this table but within our framework. Then we attempt to create RecWizard pipelines for those modules. 
\begin{enumerate}
    \item For the models~\citep{li2018towards,wang2022towards} which explicitly mentioned how to combine the outputs from both recommender modules and generator modules, we follow those instructions to build pipelines.
    \item  Nevertheless, in the case of other models~\cite{pugazhenthi2022improving,chen2019towards, he2023large}, which evaluate modules separately, we have attempted to create our own pipeline to facilitate user interface. This also highlights the current situation where the pipeline-level design and evaluation are under-emphasized.
\end{enumerate}

\xhdr{Limitations.} We discuss two major limitations of our current trained models in RecWizard framework. 
\begin{enumerate}
    \item We have not conducted a comprehensive evaluation of all trained models within a unified benchmark, as the primary focus of this work is to introduce the new features of the RecWizard toolkit. We are leaving the benchmarking of CRS with RecWizard to future research.
    \item All the models are trained using strategies similar to the original source codes, rather than a unified training strategy, as RecWizard does not offer  standardized training APIs. This decision is based on the principle that a research toolkit should allow flexibility for various training strategies. In terms of training, the modules can be treated as equivalent to regular PyTorch models. Therefore, our models can be trained with HF trainer\footnote{\url{https://huggingface.co/docs/transformers/main_classes/trainer}}, native PyTorch\footnote{\url{https://pytorch.org/}}, PyTorch Lightning\footnote{\url{https://lightning.ai/}}, or similar frameworks.
\end{enumerate}

\begin{table}[]
\scriptsize
\begin{tabular}{llll}
\toprule
\textbf{Year} &  \textbf{CRS} & \textbf{Recommender Module}                  & \textbf{Generator Module}             \\ \midrule
2018 & ReDIAL        & AutoRec + Sentiment Analysis & RNN based             \\
2019 & KBRD       & RGCN + Entity Attention      & Transformer based     \\
2020 & KGSF      & RGCN                         & Transformer           \\
2022 & UniCRS      & GNN+Prompt+GPT2              & Prompt+GPT2           \\
2023 & LLMs     & ChatGPT (with Prompt)        & ChatGPT (with Prompt) \\ \bottomrule
\end{tabular}
\caption{Current supported CRS Models in RecWizard. We provide some previous representative models~\citep{li2018towards,chen2019towards,pugazhenthi2022improving,wang2022towards} and recent LLM-based CRS~\citep{openai2022chatgpt,he2023large}.} 
\label{tab:resource}
\end{table}

\section{More Details about Toolkit Use Cases}

\subsection{An Example for Using Existing Models}

Taking \texttt{\textbf{ExpansionPipeline}} as an example, we can build a ChatGPT-based expansion model with the classic AutoRec~\citep{li2018towards} module in a few lines.

\begin{enumerate}
    \item \textbf{Setup the pipeline:} We load a RecWizard CRS model with just a few lines of code.
\begin{lstlisting}[language=python]
  from recwizard import ExpansionConfig
  from recwizard import ExpansionPipeline
  from recwizard import ChatgptGen, RedialRec

  pipeline = ExpansionPipeline(
    config=ExpansionConfig(),
    rec_module=RedialRec \
      .from_pretrained('recwizard/redial-rec'),
    gen_module=ChatgptGen \
      .from_pretrained('recwizard/chatgpt-expansion')
  )
\end{lstlisting}

    \item \textbf{Format the input:} We would want to format our conversation history like this before passing it to the pipeline:
\begin{lstlisting}[language=python]
  context = "<sep>".join([
    "User: Hello!",
    "System: Hello, I have some movie ideas for you. Have you watched the movie <entity>Forever My Girl (2018)</entity> ?",
    "User: Looking for movies in the comedy category. I like Adam Sandler movies like <entity>Billy Madison (1995)</entity> Oh no is that good?"
  ])
\end{lstlisting}

    \item \textbf{Get Response}: We would feed this conversation context and get responses soon.
\begin{lstlisting}[language=python]
  print(pipeline.response(context))
  # "If you're a fan of Adam Sandler, you might enjoy "50 First Dates (2004)" which is a romantic comedy featuring him and Drew Barrymore. \n Another one is "Big Daddy (1999)" where Sandler's character adopts a kid to impress his girlfriend, but things don't go as planned. \n Both movies have Sandler's signature humor and are quite entertaining."
\end{lstlisting}

\end{enumerate}

\subsection{A Coding Template for New Models}
\label{app:new_model}

We offer a coding template below for adding new modules and pipelines, making it easy to implement. Additionally, all the mentioned classes benefit from the HF \textbf{\texttt{save\_pretrained}} and \textbf{\texttt{push\_to\_hub}} features, ensuring they are modular and portable.

\begin{lstlisting}[language=python]
  # MODULE LEVEL
  class NewConfig(recwizard.BaseConfig):
      def __init__(self, ...):
        # specify the arguments
  class NewTokenizer(recwizard.BaseTokenizer):
      def __init__(self, entity2id, tokenizers, ...):
        # build an augmented tokenizer for words and labeled entities from textual inputs
  class NewModule(recwizard.BaseModule):
      def __init__(self, ...):
        # build it similar to HF PreTrainedModel
      def forward(self, tensor_inputs, labels):
        # define the flow of tensors through the module
      def response(self, raw_inputs, tokenizer):
        # define the raw text inputs and outputs based on forward method

  # PIPELINE LEVEL
  class NewPipelineConfig(recwizard.BaseConfig):
      def __init__(self, ...):
        # specify the arguments
  class NewPipeline(recwizard.BasePipeline):
      def __init__(self, config, rec_module, rec_tokenizer, gen_module, gen_tokenizer):
        # build it similar to HF PreTrainedModel
      def response(self, raw_inputs):
        # define the way to call `rec_module.response` and `gen_module.response` and manage the results
\end{lstlisting}



        


\section{More Examples about User Interface}
\begin{enumerate}
    \item \textbf{Basic functions} The interactive interface starts with pipeline selection, which allows users to select different pipelines that is built in the package (Figure~\ref{fig:model-selection}). Users can send new message to the system by typing in the text box and sending the message with the ``send'' button or with a carriage return. The system will return the response with the movie names linked to external links. Users can also choose to stop generation or refresh the chat session.
    \item \textbf{Run-time options}: In the sidebar, the user can specify different keyword arguments for the response method of recommenders and generators. Options like GPT~\citep{openai2022chatgpt} model selection are also supported. The user can also download the chat history clicking a button.
    \item \textbf{DEBUG mode}: In this mode, you can see the response time for each method decorated with \textbf{\texttt{@monitor}} (Figure~\ref{fig:model-monitoring}), shown either in a timeline or in a network graph. You can also look at the inputs and outputs for these methods, which well facilitates the debugging process.
\end{enumerate}
\begin{center}
  \centering
    \includegraphics[width=0.9\columnwidth]{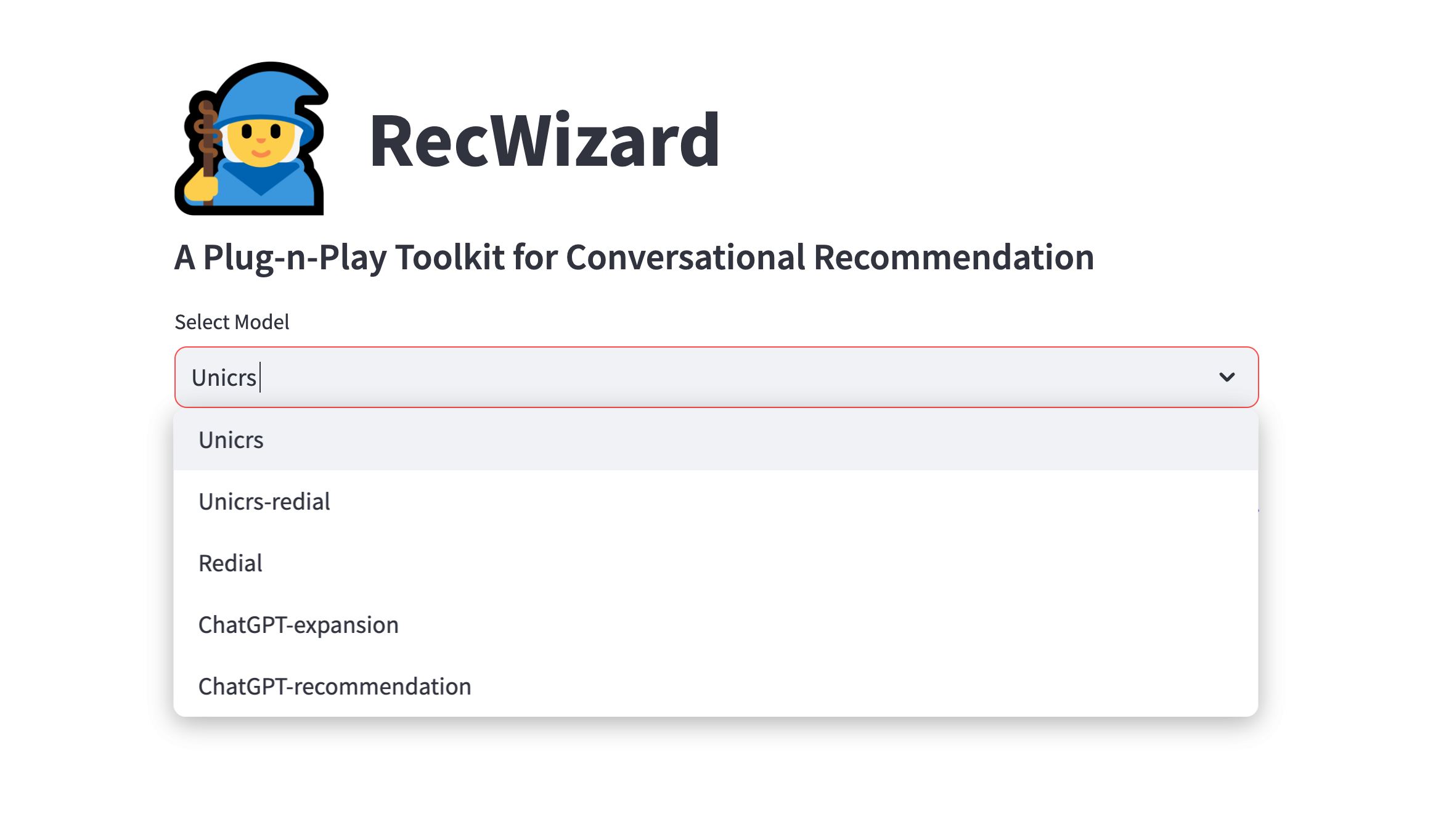}
  \captionof{figure}{Model selection in user interface.}
  \label{fig:model-selection}
\end{center}

\begin{center}
  \centering
    \includegraphics[width=0.9\columnwidth]{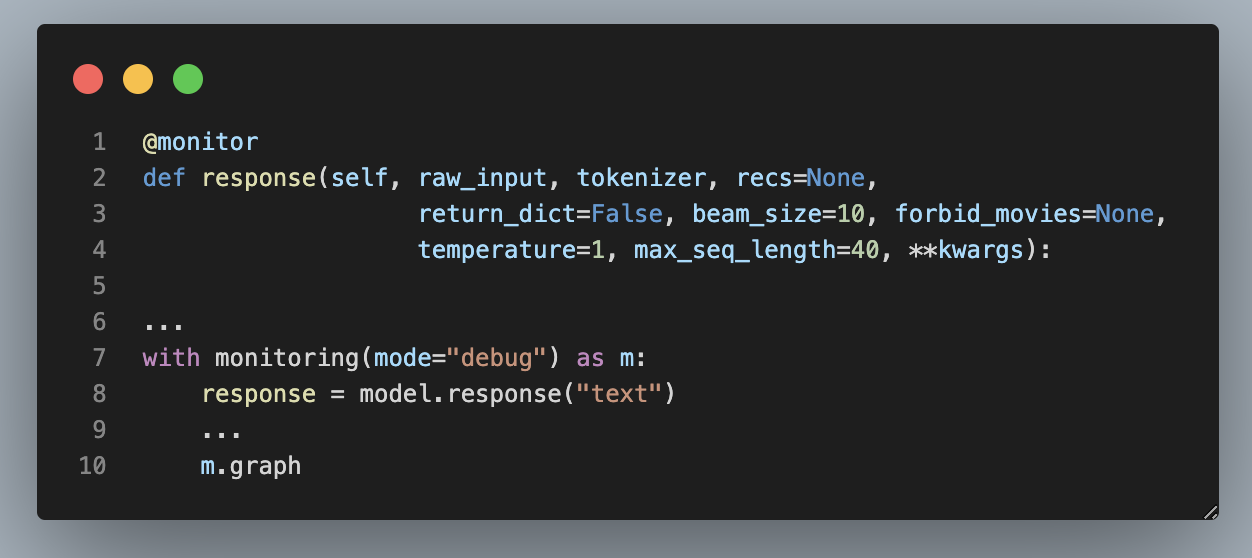}
  \captionof{figure}{Monitoring example.}
  \label{fig:model-monitoring}
\end{center}

\end{document}